\documentstyle[sprocl]{article}

\def\be{\begin{equation}}
\def\ee{\end{equation}}
\def\bea{\begin{eqnarray}}
\def\eea{\end{eqnarray}}

\newcommand{\slashed}[1]{\rlap{$#1$}/}

\begin{document}

\title{THE ASYMPTOTIC BEHAVIOR OF THE
$\pi^0 \gamma^\star \gamma$ TRANSITION}

\author{DUBRAVKO KLABU\v{C}AR}

\address{Physics Department, Faculty of Science,
        POB 162, University of Zagreb, Croatia}

\author{DALIBOR KEKEZ}

\address{Rudjer Bo\v{s}kovi\'{c} Institute,
         POB 1016, 10001 Zagreb, Croatia}


\maketitle\abstracts{ 
It is discussed how various {\it Ans\" atze} for the dressed 
quark-photon ($qq\gamma$) vertices $\Gamma^\mu(q,q^\prime)$
influence the asymptotics of the $\gamma^\star \gamma \to \pi^0$
transition form factor.
}

It has been shown \cite{KeKl3} that the Schwinger-Dyson (SD) approach 
to physics of quarks and gluons leads to the
$\gamma^\star(k)\gamma(k^\prime)\to\pi^0(p)$ transition form factor
$T_{\pi^0}(k^2,{k^\prime}^2)$ which has the asymptotic momentum dependence
\begin{equation}
 T_{\pi^0}(-Q^2,0) =  \frac{{\cal K}}{Q^2} \,
\qquad ({\cal K} \to {\rm const} \,\, {\rm as \,\,} Q^2 \to \infty),
\label{largeQ2}
\end{equation}
for large spacelike $k^2 = -Q^2 <0$.
This is consistent with the data \cite{gronberg98} at the presently 
largest accessible $Q^2$ and in agreement (up to the precise value of
${\cal K}$) with perturbative QCD where~\cite{BrodskyLepage}
${\cal K}=2 f_\pi$, operator product expansion (OPE) where~\cite{manohar90}
${\cal K}=4 f_\pi/3$, and QCD sum rules where~\cite{Radyushkin+Rusk3}
${\cal K}\approx 1.6 f_\pi$.

The SD approach, where the quark propagators
$S(q) = [ A(q^2)\slashed{q} - B(q^2) ]^{-1}$ are dynamically dressed,
also requires consistently dressed quark-photon ($qq\gamma$) vertices
$\Gamma^\mu(q,q^\prime)$ in order to satisfy the vector Ward-Takahashi 
identity (WTI).
[In addition to using for the pseudoscalar vertex the quark-antiquark
pion Bethe-Salpeter (BS) bound-state vertex $\Gamma_{\pi^0}(q,p)$,
this defines {\it generalized impulse approximation} (GIA).]
Even just approximately adequate SD solutions for
$\Gamma^\mu(q,q^\prime)$ are not yet available, and in practice the
more or less realistic WTI-satisfying {\it Ans\" atze} still must be
used.

The topic of this writeup is the dependence of the asymptotic 
coefficient ${\cal K}$ on the choice of the {\it Ansatz} for 
$\Gamma^\mu(q,q^\prime)$. This topic needs clarification, since 
Ref. \cite{TandyDub99} expressed a slight disagreement about one 
detail in what we found \cite{KeKl3}.
(The rest of the material of my talk is amply covered
in Refs. \cite{KeKl3,KeBiKl98,KlKe2,KlKe4}.)

Ref. \cite{KeKl3} showed that the SD approach predicts
${\cal K}=4 f_\pi/3$ (the
same as OPE~\cite{manohar90}) for {\it all} $qq\gamma$ vertices
$\Gamma^\mu(q^\prime,q)$ which go into the
bare one ($\gamma^\mu$) even if just {\it one} of the squared
momenta $q^2$ or $q^{\prime 2}$ becomes infinite. This was
illustrated on the examples of the Curtis-Pennington (CP)
vertex \cite{CP90} $\Gamma^\mu_{CP}$ and the modified
Curtis-Pennington (mCP) vertex $\Gamma^\mu_{mCP}$ \cite{KeKl3}.
(For the latter, $T_{\pi^0}(-Q^2,0)$ was calculated also for
finite values of $Q^2$.) Both the CP and mCP vertices are
multiplicatively renormalizable, so that
our result \cite{KeKl3} on the asymptotic behavior of
$T_{\pi^0}(-Q^2,0)$ subsequently received further support
from Ref. \cite{RobertsDubr}. This reference generalized 
our derivation \cite{KeKl3}
by taking into account renormalization explicitly,
showing that the asymptotics of Ref.~\cite{KeKl3}
with ${\cal K}=4 f_\pi/3$ must hold
for any $qq\gamma$ vertex which is
consistent with multiplicative renormalizability.

However, Ref. \cite{KeKl3} also showed that the usage of the ``minimal"
WTI-satisfying $qq\gamma$ vertex $\Gamma^\mu_{BC}$, namely the
Ball-Chiu (BC) one, leads to the asymptotic coefficient
${\cal K}=4 {\widetilde f}_\pi/3$, where ${\widetilde f}_\pi$ is the
quantity given by the same Mandelstam-formalism expression as the pion
decay constant $f_\pi$, except that the integrand is modified by the
factor $[1+A(q^2)]^2/4$. (In the case of our solutions \cite{KeBiKl98},
this gives ${\widetilde f}_\pi = 1.334 f_\pi = 124$ MeV.)
Note that the arguments of Ref. \cite{RobertsDubr} do not preclude
the change $f_\pi \to {\widetilde f}_\pi$,
since the BC vertex is {\it not} consistent with
multiplicative renormalizability \cite{CP90}.
This modification of ${\cal K}$ is caused by the different asymptotic
behavior of the BC vertex, which tends to the bare vertex,
$\Gamma^\mu_{BC}(q^\prime,q)\to\gamma^\mu$, only when
the squared momenta in {\it both} fermion legs tend to infinity,
$q^{\prime 2},q^2\to \pm \infty$.
The origin of the factor
$(1/2)^2[1+A([q+p/2]^2)][1+A([q-p/2]^2)] \approx [1+A(q^2)]^2/4$
modifying the integrand
when $\Gamma^\mu(q^\prime,q)=\Gamma^\mu_{BC}(q^\prime,q)$ is then
clear: $T_{\pi^0}(k^2,{k^\prime}^2)$ is
extracted from the tensor amplitude $T_{\pi^0}^{\mu\nu}(k,k^\prime)$
for the GIA triangle diagram,
\begin{displaymath}
        T_{\pi^0}^{\mu\nu}(k,k^\prime) \propto
        \int\frac{d^4q}{(2\pi)^4} \mbox{\rm tr} \{
        \Gamma^\mu(q-\frac{p}{2},k+q-\frac{p}{2})
        S(k+q-\frac{p}{2})
\end{displaymath}
\begin{equation}
          \qquad
        \times
        \Gamma^\nu(k+q-\frac{p}{2},q+\frac{p}{2})
        S(q+{p}/{2})\Gamma_{\pi^0}(q,p)S(q-{p}/{2}) \}
        +
        (k\leftrightarrow k^\prime,\mu\leftrightarrow\nu) \, ,
\label{Tmunu(2)}
\end{equation}
and since all quark loop momenta $q$ contribute, the small values of
$(q\pm p/2)^2\approx q^2$ in one quark leg will prevent the BC vertex
$\Gamma^\mu_{BC}(q^\prime,q)$ from reducing always to the bare
$\gamma^\mu$-vertex, even when a hard virtual photon momentum $k^2 = -Q^2$
makes ``bare" the other fermion leg in the vertices
$\Gamma^\mu(q^\prime,q)=\Gamma^\mu_{BC}(q^\prime,q)$.

However, this result on the asymptotics of $T_{\pi^0}(-Q^2,0)$ when
using the BC vertex, caused some controversy since Ref. \cite{TandyDub99}
claimed that even for the BC vertex
${\cal K}=4 f_\pi/3$, {\it i.e.}, that no modification occurs for the BC
vertex due to one soft quark leg. The argument of Ref. \cite{TandyDub99}
(see its Sec. {\it 4.})
is that there are in fact no soft legs in the
$qq\gamma$ vertices when $Q^2$ becomes very large. The on-shell condition
for the pion and one photon,  $(p - k)^2 = M_\pi^2 - 2 p\cdot k -
Q^2 \approx - 2 p\cdot k - Q^2 = {k^\prime}^2 = 0$, and $k^2 = - Q^2$,
are used to argue that the pion momentum $p$ has components which must
scale like $k$ and thus like $Q$. Then,
$A([q\pm p/2]^2)=A(q^2 \pm q\cdot p + M_\pi^2)$ would tend to
1 as $Q^2\to \infty$ even for very soft loop momenta $q$, just because
of $p \sim Q$, causing $\Gamma^\mu_{BC} \to \gamma^\mu$.

We will now demonstrate that this argument does not hold.
The very fact that the size of the
{\it components} is invoked makes the argument suspect,
because it is a frame-dependent statement.
The argument of Ref. \cite{TandyDub99} relies on
working in a Lorentz frame such as the one where $k=(0,0,0,\sqrt{Q^2})$,
$k^\prime = (E_\pi,0,0,-E_\pi)$, $p = (E_\pi,0,0,\sqrt{Q^2}-E_\pi)$, and
$E_\pi = (Q^2+M_\pi^2)/2 \sqrt{Q^2}$. Even if one sticks just to that
choice in one's calculation, one can expect persistent soft contributions
because of those soft loop momenta $q$ which are also perpendicular to $p$
so that $p\cdot q =0$. However, the shortest and clearest demonstration
that, at least in this application, $q\cdot p$ cannot be hard if $q$ is
soft, is noting that one can make a Lorentz transformation to the pion
rest frame. In this case it is the boost transformation along
the $z$-axis and with the parameter $\beta = (Q^2-M_\pi^2)/(Q^2+M_\pi^2)$.
In that frame, $k=(M_\pi-E_\gamma,0,0,E_\gamma)$ and
$k^\prime=(E_\gamma,0,0,-E_\gamma)$,
with $E_\gamma=(Q^2+M_\pi^2)/2 M_\pi$, whereas $p=(M_\pi,0,0,0)$,
making it clear that for the light pion, $A([q\pm p/2]^2)$ is
approximated well by $A(q^2)$ and not by $A(\pm q\cdot p)$ which
allegedly \cite{TandyDub99} would be 1.

We want to make clear that we of course give precedence to
the value ${\cal K}=4 f_\pi/3$ for the asymptotic coefficient
as the one having the more fundamental meaning, resulting
from the $qq\gamma$ vertices such as the CP or mCP ones,
which have properties closer to the true vertex solution,
such as being renormalizable.
Also indicative is the asymptotics found by Ref. \cite{KeKl3}
for the case when  both photons are off-shell, $k^2 = - Q^2 << 0$ and
$k^{\prime 2} = - Q^{\prime 2} \leq 0$:
\begin{equation}
T_{\pi^0}(-Q^2, -Q^{\prime 2}) = \frac{4}{3} \,
                                \frac{f_\pi}{Q^2 + {Q^{\prime}}^2} \, .
                                \label{baregasgas}
\end{equation}
This is found for the $qq\gamma$ vertices which reduce to the bare
$\gamma^\mu$ as soon as just one of the quark legs is hard,
while the usage of the BC vertex again modifies
this result by the substitution $f_\pi \to {\widetilde f}_\pi$.
Eq.~(\ref{baregasgas}) agrees with the leading term of the OPE result
derived by Novikov $et \,\, al.$ \cite{novikov+al84} for the special
case $Q^2=Q^{\prime 2}$. The distribution-amplitude-dependence
of the pQCD approach cancels out for that symmetric case,
so that $T_{\pi^0}(-Q^2, -Q^{\prime 2})$ in this approach ({\it e.g.},
see \cite{Kess+Ong93}), in the limit $Q^2={Q^{\prime}}^2\to\infty$,
exactly agrees with both our Eq. (\ref{baregasgas}) and Ref.
\cite{novikov+al84}. Therefore, for that symmetric case,
we should have even the precise agreement of the coefficients
irrespective of the description of the pion internal structure
encoded in the distribution amplitude. Obviously, this favors
the $qq\gamma$ vertices which reduce to the bare $\gamma^\mu$ 
as soon as one of the quark legs is hard,
over the BC vertex, and ${\cal K}=4 f_\pi/3$ over
${\cal K}=4 {\widetilde f}_\pi/3$. However, the BC vertex,
which is the simplest WTI-preserving vertex and has been
the one most widely used in phenomenological applications,
may anyway be the one which is more accurate not only for the
presently accessible $Q^2$, but also for much larger values
before starting to fail. For that reason it is important to
understand the asymptotic behavior to which the BC vertex leads.

\section*{Acknowledgments}
D.K. wants to thank David Blaschke (who, as the workshop coordinator
secured support and made D.K.'s participation possible), Fritjof Karsch
and Craig Roberts for organizing International Workshop on
Understanding Deconfinement in QCD, ECT* Trento, Italy,
March 01 - 13, 1999, where this work was initiated, and
to P. Tandy and P. Maris for the valuable discussions
in connection with this.

\end{document}